# The "Machinery" of Biocomplexity: understanding non-optimal architectures in biological systems

Bradly Alicea[1]
bradly.alicea@outlook.com

## ABSTRACT

One popular assumption regarding biological systems is that traits have evolved to be optimized with respect to function. This is a standard goal in evolutionary computation, and while not always embraced in the biological sciences, is an underlying assumption of what happens when fitness is maximized. The implication of this is that a signaling pathway or phylogeny should show evidence of minimizing the number of steps required to produce a biochemical product or phenotypic adaptation. In this paper, it will be shown that a principle of "maximum intermediate steps" may also characterize complex biological systems, especially those in which extreme historical contingency or a combination of mutation and recombination are key features. The contribution to existing literature is two-fold: demonstrating both the potential for non-optimality in engineered systems with "lifelike" attributes, and the underpinnings of non-optimality in naturalistic contexts.

This will be demonstrated by using the Rube Goldberg Machine (RGM) analogy. Mechanical RGMs will be introduced, and their relationship to conceptual biological RGMs. Exemplars of these biological RGMs and their evolution (e.g. introduction of mutations and recombination-like inversions) will be demonstrated using block diagrams and interconnections with complex networks (called convolution architectures). The conceptual biological RGM will then be mapped to an artificial vascular system, which can be modeled using microfluidic-like structures. Theoretical expectations will be presented, particularly regarding whether or not maximum intermediate steps equates to the rescue or reuse of traits compromised by previous mutations or inversions. Considerations for future work and applications will then be discussed, including the incorporation of such convolution architectures into complex networks.

## Introduction

In the early 20th century, a cartoonist named Rube Goldberg drew a series of cartoons that featured "absurd" machines. An archetypical Rube Goldberg Machine (RGM) performs a simple task such as flipping a light switch using as many intermediate steps as possible. These intermediate steps are linked together in a serial fashion, so that each preceding step triggers all subsequent steps. The many and varied creations of Rube Goldberg have also inspired engineering tournaments that treat such designs as a curiosity [1] (see Figure 1). The winning creations are judged more in terms of creative value rather than their efficiency.

In the realm of biological systems and evolutionary biology, however, we find that functioning systems are often not optimal in terms of their form and/or function. The principle of

---
[1] Orthogonal Research, Champaign, IL. USA

"maximum intermediate steps" seems to fit a number of empirical observations of the general structure [2] and function [3,4] of biological systems. This condition arises as a consequence of complex traits being built from a series of responses to challenges distributed across an organism's evolutionary history rather than arising *de novo*[2]. Biological RGMs (and by extension convolution architectures) are also characterized by the use of unconventional pathways to accomplish a function, such as the appendix in mammalian immune systems. While the appendix appears to be a vestigial organ, it may actually serve as a functional component of lesser importance [3, 4]. Biological traits that function as processes, such as inversion of the dorsoventral axis in vertebrate development [6] and the evolution of derivative sensors in signal transduction pathways [7], are particularly well suited to the RGM model.

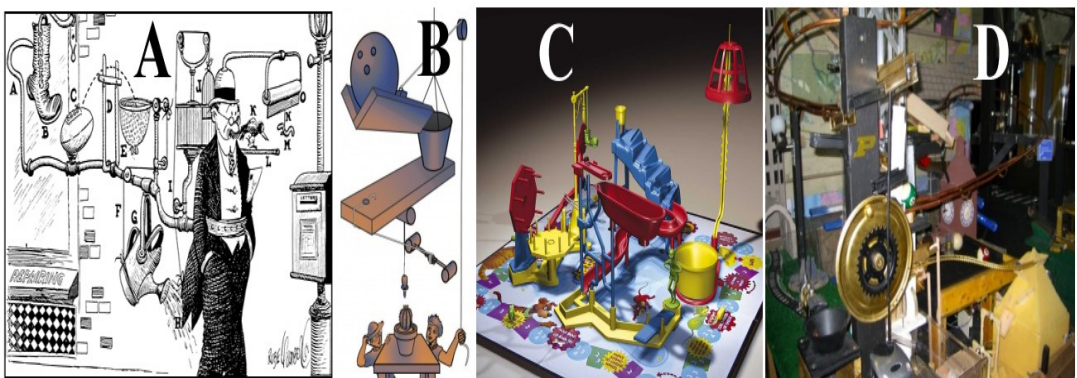

**Figure 1.** Examples of Rube Goldberg Machines (RGMs). From left: A) a 13[th] order mechanical RGM (steps A-O, in cartoon form). B) cartoon of a 4[th] order mechanical RGM, C) a photograph of the boardgame "Mousetrap". D) photo of a mechanical RGM from the national collegiate tournament at Purdue University [8].

In this paper, the potential of RGMs as a model for evolution and complex, multiscale phenomena in biological systems will be explored. The use of RGMs stand in contrast to most theoretical models of evolutionary biology and optimization, which assume that natural selection optimizes traits and their function over time [9, 10]. However, the working hypothesis in this paper will be that for some traits, the presence of a large number of intermediate steps can actually maximize fitness. The first section of this paper will lay out how RGMs can be mapped to biocomplexity. The second section will detail how intermediate steps in a process relate to the control, function, and robustness of a biological trait. The final section will extend the metaphor to more formal mathematical constructs such as Markovian dynamics [11].

**Biological RGMs in Evolution**
Applying biological RGMs to evolution allows us to question whether or not natural selection favors an optimal architecture for any particular trait. Examples such as the recurrent morphogenesis in the vertebrate retina [2, 12], inversions in development that give rise to major shifts in body plan [13], and the evolution of signaling pathways [14] all show some evidence of a maximization scheme with regard to the number of discrete steps in a biological process. The maximization of steps in a process can potentially maximize fitness in a number of ways. The most likely way for this to be so is by one of more steps in the biological RGM acting as a bridge or pivot to other traits. This is related to the degree of serial function ascribed to a biological RGM, which will be discussed in more detail later.

---

[2] Ernst Haeckel [5] used the idea of dysteleology to describe evolution with no overarching goal, which has been recapitulated in modern contexts as the phrase "incompetent design".

Biological RGMs can also be compared to the hypercube approach used for modeling evolvability in RNA phenotypes [15]. A network topology of phenotypes, each node defined by a unique *n*-tuple, is used to characterize the mutational distance between all possible phenotypes. This allows us to understand how many mutations are required to transform a phenotype from one state to another. Adami [16] has demonstrated that highly specific lock-and-key receptor systems can evolve gradually using a small number of steps. In the method presented here, this work is taken a few steps further by showing that historical contingency[3] in conjunction with discrete, localized changes can produce complex but functional traits without the requirement of an overarching goal.

**Biological RGMs in Multiscale Processes**
Relative to evolutionary processes, multiscale processes are a mystery. However, biologicial RGMs can also provide insight into these phenomena. Multiscale processes usually involve the mapping of inputs at one scale (e.g. gene expression) to outputs at another scale (e.g. phenotype). In some cases, this relationship can be highly convoluted. For example, processes that occur at different scales of organization can also occur at different rates. However, research from the computational science community suggests that continuum equations can be distilled from representations of differential granularity at other scales [20].

**RGMs as Biological Heuristics**
In order to better understand exactly what RGMs do and how they might get there in evolution and development, we must create a heuristic for the biology of specific traits and processes that focuses on their discrete dynamical properties [21]. What remains is a model that shares similarities with a mechanical RGM that will demonstrate how the number of steps between an input and an output. The inputs could represent sensory information or chemical energy, while the outputs could represent phenomena such as a muscle contraction or speech. The proposed conventions for modeling biological RGMs are presented in the Methods section.

**Examples of Biological RGM Function**
Two hypothetical examples of biological RGM function are shown in Figure 2. The schematics in Frames A-C demonstrate the sequence of events that might occur as a biological RGM self-organizes and maintains itself in response to perturbations. Mutation/Co-option (Scenario #1) involves a mutation in A so that it no longer communicates with B, and a mutation in the connectivity between B and C which confers bidirectionality. To overcome the lack of connectivity between A and B, a new element (D) evolves to restore connectivity. Inversion (Scenario #2) involves an inversion of elements B and C so that an obstacle exists between A and C (shown in Frame B). To alleviate this bottleneck and produce a viable output without additional mutation, element D and E are added with appropriate connectivity.

---

[3] Historical contingency [for examples, see 17-19] means that all future options of a system are determined by events that occurred previously. In the context of a directed graph, this assumes a strict hierarchy which confines evolved phenomena to a search subspace.

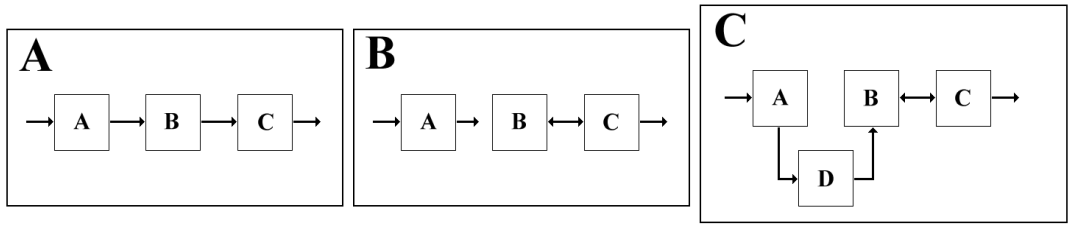

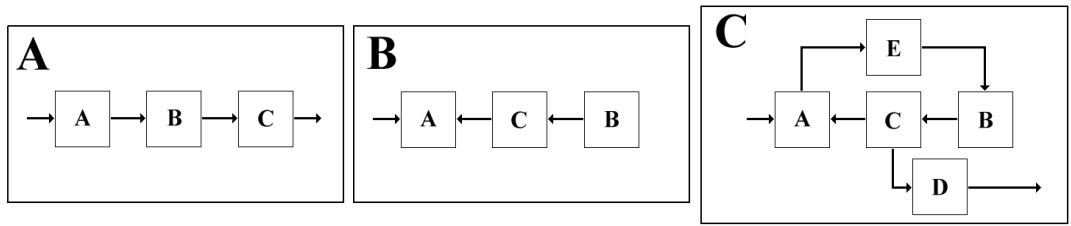

**Figure 2.** Two hypothetical scenarios for the formation of a biological RGM. Scenario #1 (mutation/co-option) is shown in Frames A-C (top). Briefly, the introduction of two separate mutations in Frame B require the addition of an extra element (D) shown in Frame C. Scenario #2 (inversion) is shown in Frames A-C (bottom). Briefly, an inversion of elements in Frame B requires two additional elements (D and E) shown in Frame C.

While both of these examples are evolutionary in nature, each typify how a biological RGM is built as needed on top of older structures. A scenario specifically related to biocomplexity is shown in Figure 3. The multi-scale RGM involves several subsets of elements, which can be arranged singularly or in parallel and represent a distinct hierarchical level of biological organization. An example of this parallelism can be seen with element group $E_{m,n}$. Subset E includes five identical elements placed in parallel, each receiving connections from group $D_{m,n}$ (containing 3 elements) and sending connections to group $F_{m,n}$ (also containing 3 elements). These elements are fully connected, which means that variability and convolution in RGM function is maximized between scales. More details about how a convoluted output is determined can be found in the Methods section.

**Model order and consequences on function.** The lower portion of Figure 2 showing inversion in a biological RGM can be used to demonstrate the order of a given model. A zero-order model would be characterized by one element with an input and output. The initial condition shown in Frame A represents a second-order model, with three elements connected serially. A portion of this second-order model is then inverted in Frame B, which violates the "domino effect" of serial flow. The result, shown in Frame C, is the addition of two elements. One element (D) reestablishes an output, while the other element (E) reestablishes a connection between element A and element B and thus serial flow. The feedback from element C to element A is an unintentional consequence of the original inversion, and may require additional elements for purposes of self-regulation.

**Formation and Maintenance of RGM Function.** Conditions for the formation and maintenance of biological RGNs may be present in both evolutionary and multi-scale contexts. While examples of evolutionary and multi-scale causal factors are provided separately, it is important to remember that the survival of an organism across evolution is dependent upon good solutions to multi-scale complexity and vice versa.

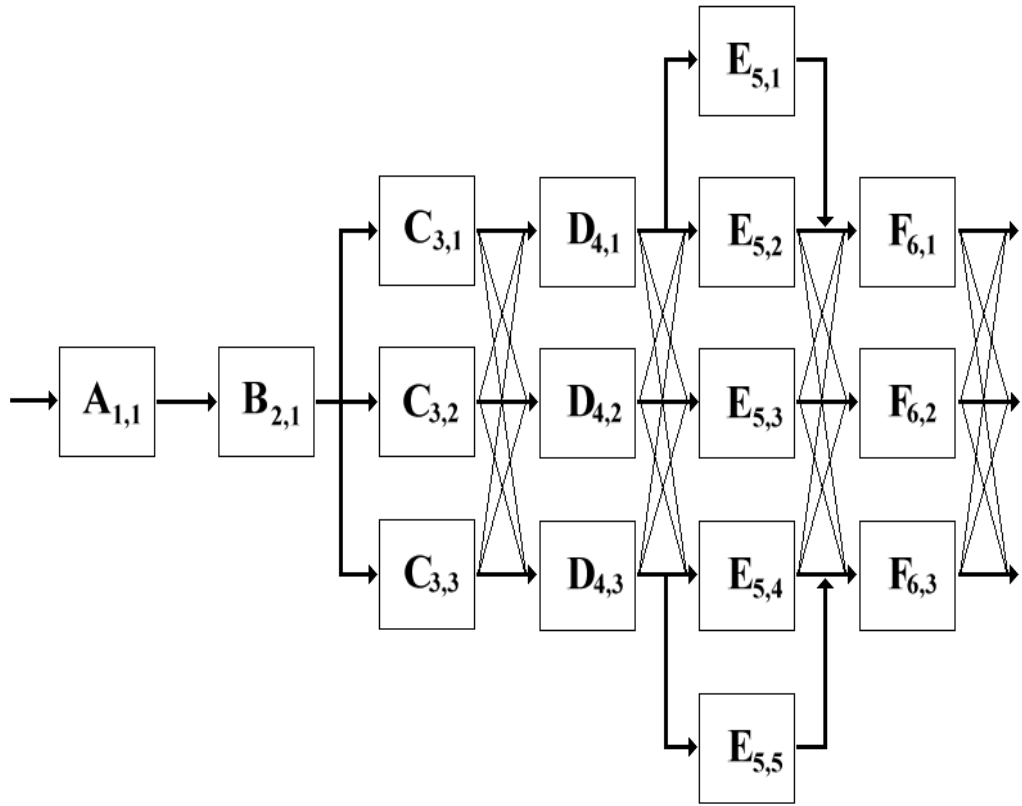

**Figure 3.** A hypothetical biological RGM representing a multi-scale relationship (e.g. genotype to phenotype). Each set of elements ($A_{n,m}$-$F_{n,m}$) represents the number of elements at each scale. In general, the greater number of elements existing at each scale, the greater potential for convolution in the output on the right.

In evolutionary contexts, a biological RGM might result from a variety of causal factors. One of these causal factors involves the different components of a biological RGM belonging to different developmental modules [see 15]. Developmental modules such as different body segments may grow and mature at different rates, so the optimization in function of such a trait would involve a massive and possible lethal changes reorganization of the organism's anatomy and development. The multi-scale case involves the mapping from genes and phenotype. In multi-scale contexts, one causal factor might involve variable biochemical kinetics between scales and other self-regulatory factors. Given these challenges, maintaining solutions for mapping a genotype to a phenotype may only be allowable through the use of complex, dynamic mechanisms with many redundant components. This would be particularly true of an undirected system that adapts to its environment.

**Microfluidic-inspired RGM Modeling**

Now that a computational representation for understanding biological RGMs has been made familiar, a microfluidic-inspired physical model can be used to understand how a biological RGM might arise. Microfluidics is used the field of Biological MEMS (microelectromechanical systems) and Bioengineering to model flows at micron ($10^{-7}$) to nanometer ($10^{-9}$) scales [22]. The model shown here is a quasi-evolutionary model that does not require a fitness function, but does provide a window into evolutionary processes. On a microfluidic chip, the default state of a channel is a laminar flow. This provides a baseline which is analogous to the maintenance of constant shear stress in healthy blood vessels and other vasculature [23]. Orthogonal, diagonal, and cataract-like obstacles can then be added along the walls of the channel to make the flow selectively turbulent (Figure 4 and Table 1).

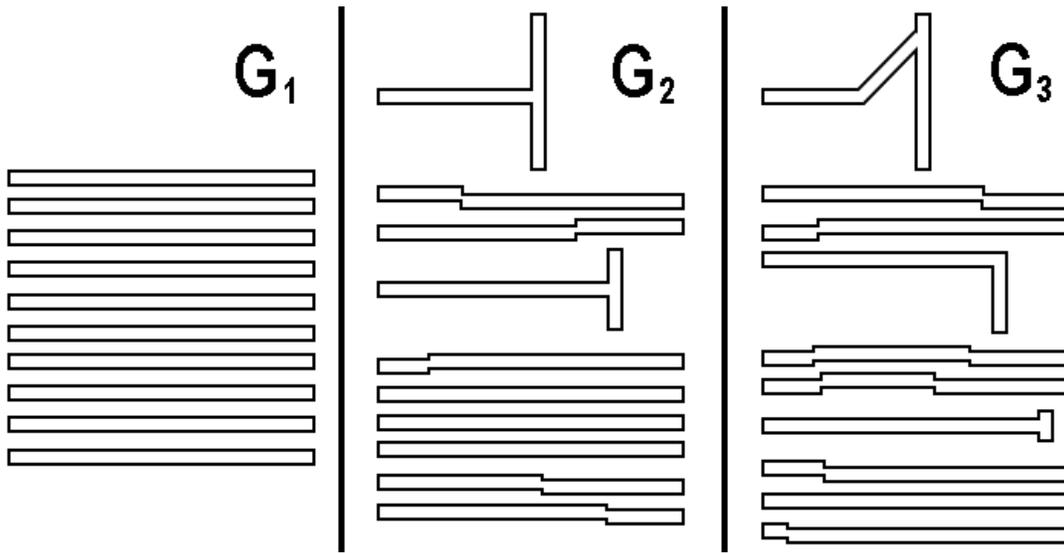

**Figure 4.** Microfluidic-like structures that mimic fluid flows in vasculature. Generation 1 (G1 – left) shows an initial population of 10 channels (arranged from top to bottom). Generation 2 (G2 – middle) shows a variety of single mutation or inversion events. Generation 3 (G3 – right) shows changes that either compensate or further compromise the function of each channel.

In this set of examples, it will be assumed that a fluid moves from one end of a channel to the other. Over the course of time, obstacles that represent instances of mutation and inversion will be added to the channel. This will help to rescue laminar flow in the channel, much like shear forces might be selected for and rescues in real vasculature. The number of innovations required to overcome these randomly-introduced obstacles will be the metric of RGM-ness (Figure 5).

**Understanding Biocomplexity Using RGMs**

In mapping RGMs to observable biocomplexity, an analogical bridge needs to be made between the purely physical mechanisms of the model and biologically-specific mechanisms found in nature. In doing so, two features of biological systems need to be considered: the degree to which function within the RGM is serial, and the diversity of function. The degree of serial of function involves the extent to which components of the trait or process in question are mutually exclusive with regard to other systems. The diversity of function relates directly to promiscuity, and does not directly imply parallelism.

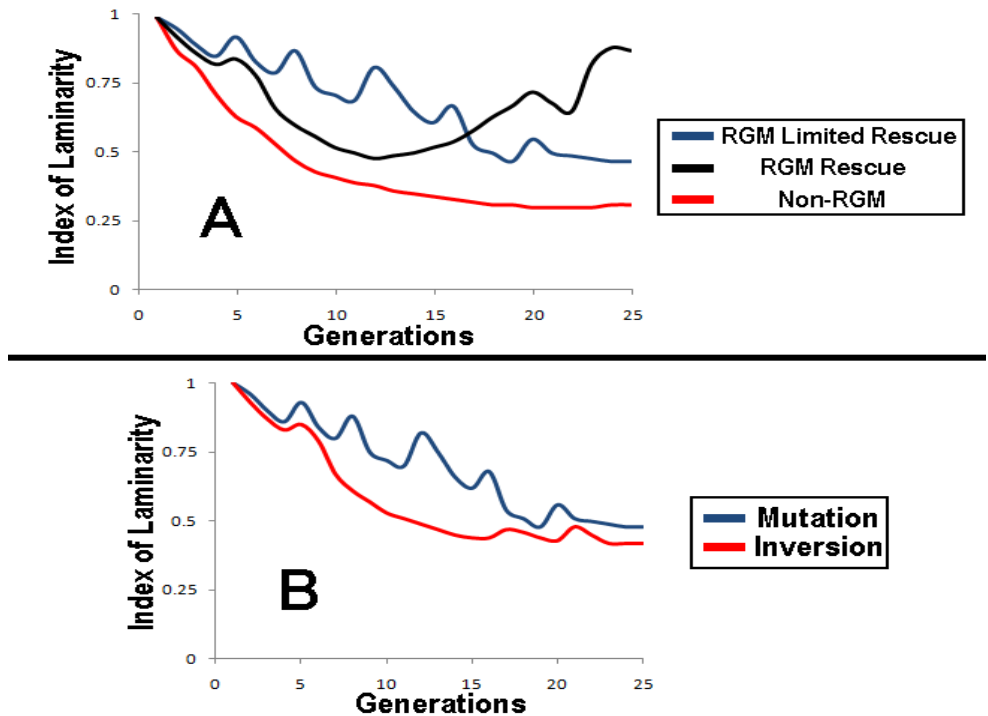

**Figure 5.** Schematics (using pseudo-data) demonstrate the dynamic behavior of the artificial vasculature when mutation or inversion is introduced. A) RGMs are thought to either rescue the phenotype almost entirely (black), or do so in a limited fashion (blue). Non-RGM behavior is shown in red. B) effects of mutations only (blue) versus inversions only (red). The mutations only condition is thought to mimic the RGM rescue condition in A, while the inversions only condition is thought to mimic the RGM limited rescue condition in A.

In the graphs of Figure 5, it is shown that a biological RGM may rescue a phenotype over evolutionary time. This condition is demonstrated by the black function in Frame A, and generally occurs by going through a relatively large number of steps. In biological systems, this is analogous to the rescue of a phenotype shown experimentally through the forced overexpression of a single mutant gene [24]. In the case of biological RGMs, however, rescue is more likely due to evolutionary consequences such as the co-option of traits or rewiring of a gene network. In the blue function of Figure 5, frame A, a condition called "limited rescue" is used to demonstrate how a trait can oscillate between restoring a laminar flow and being impeded by changes introduced in evolution. This is analogous to how a functional phenotype might be rescued many times over in the course of evolution.

**Degree of Serial Function**
One reason why RGMs are expected to exist in a biological context is that each component is partially homogeneous with regard to function. In non-biological RGMs, the absolute serial dependence of components is similar to the well-known "domino effect", in which the behavior of previous elements has a direct bearing on the behavior of the currently active element. In evolutionary terms, this might be characterized as extreme historical contingency.

**Table 1.** Changes by generation for each microfluidic-like channel among the 10-member population shown in Figure 4.

| $G_1$ | $G_2$ | $G_3$ |
|---|---|---|
| None | Inversion (orthogonal) | Inversion (diagonal) |
| None | Mutation (left end) | Mutation (middle) |
| None | Mutation (right end) | Mutation (middle) |
| None | Inversion (orthogonal) | Mutation (upper end) |
| None | Mutation (left end) | Mutation (right end) |
| None | None | Mutation (middle) |
| None | None | Inversion (orthogonal) |
| None | None | Mutation (left end) |
| None | Mutation (right end) | Mutation (left end) |
| None | Mutation (right end) | Mutation (middle) |

**Diversity of Function**

The diversity of function essentially determines the degree of serial function for each biological RGM component. This heterogeneity may be manifest in the functional promiscuity of each component. In a biological context, RGMs will also imbue the organism surrounding it with a number of properties. Of particular interest are properties relating to controllability, functionality, and robustness. To get a better appreciation of the role each of these concepts plays in the function of a biological RGM, they must be defined and placed in biological context.

**Controllability.** The controllability of a given system [25] determines how well a system can be self-regulated. In homeothermic animals, body temperature is considered controllable. In non-diabetics, blood glucose is also considered controllable. However, tumor growth in many cancers is not controllable. In light of these examples, highly serial biological RGMs seem to be state controllable [24]. State controllability must be achieved given current variable values and no direct information about the past. Related to controllability are the simple and higher-order control functions a biological RGM can perform. This capability is for RGMs that are not strictly serial, and therefore interact with other traits and subsystems.

**Functionality.** Functionality in a biological RGM can be determined in two ways. This can be thought of as two alternate viewpoints: prospective and retrospective. Determining the functionality of a biological RGM involves an understanding of how each component interacts and contributes to the output of the machine. By this criterion, each component of a biological RGM is both necessary and sufficient. On the other hand, biological RGMs might evolve by the co-option of existing components into a coherent trait.

**Robustness.** The robustness, or invariance to perturbation, exhibited by a biological RGM requires us to speculate a bit with regard to the evolutionary capacity of the RGM as a whole. In short, evolutionary capacity and robustness seem to be a consequence of a given system's functionality.

## RGMs as Dynamical Systems

Since the function of an RGM unfolds over time, particularly with a critical dependence on prior events, these structures can be considered using a formal dynamical systems approach. This critical function is governed by classical mechanics, as the function of a non-biological RGM is based on the kinematic relationship between its components. One way to characterize dynamical systems that emphasize kinematics is to use a Markovian perspective. One way in which biological RGMs could be further formalized is by using a Markov model, which characterizes the biological RGM as an ensemble of discrete entities operating in discrete time.

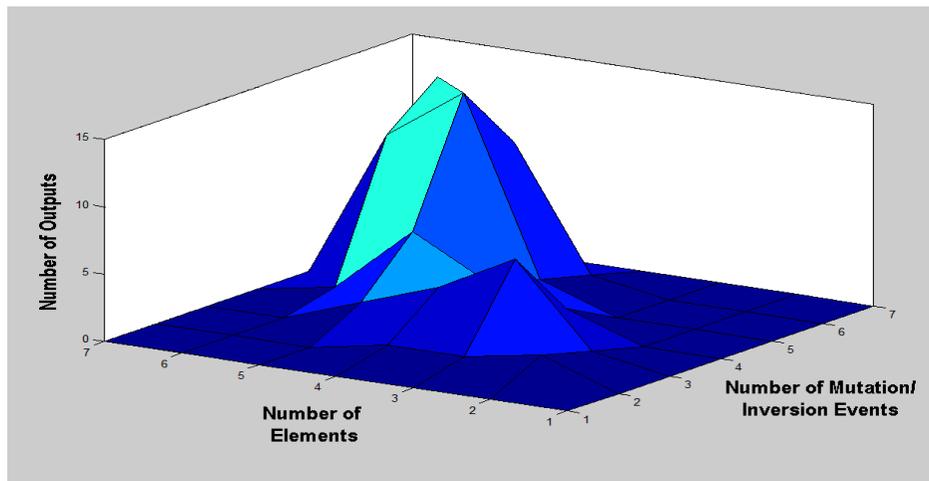

**Figure 6.** Schematic (using pseudo-data) showing the relationship between number of elements, number of outputs, and number of mutation/inversion events.

## Towards Massively Interconnected Convolution Architectures

While standalone biological RGMs can help us approximate the evolution of specific circuits and complex traits with many moving parts, we can also embed biological RGMs into complex network topologies to approximate the evolution of subtle, sub-optimal biological responses. One way to understand complexity in biological networks is to isolate simple motifs like switches and bi-fans. However, this does not fully capture the outcomes of evolutionary processes. Convolution architectures (which include biological RGMs) can demonstrate bricolage and ad-hoc formation of new mechanisms atop existing complexity. Unlike simple motifs approximated by biological RGMs, these models are intended to demonstrate how evolution can produce complex processes that operate in a sub-optimal fashion.

**Biological RGMs in Context**

By themselves, biological RGMs are capable of great convolution with respect to output. Figure 7 demonstrates how the process flow through a biological RGM can be convolved given just two mutational steps from an optimized initial condition of minimal complexity. This toy demonstration shows that the information of such architectures can be reduced to two binary matrices: connectivity and process flow. However, consider that such architectures represent a single genetic circuit (e.g. a first-order motif). How might we characterize the true complexity of an organism's genetic background relative to the evolution of this circuit?

## Initial Condition

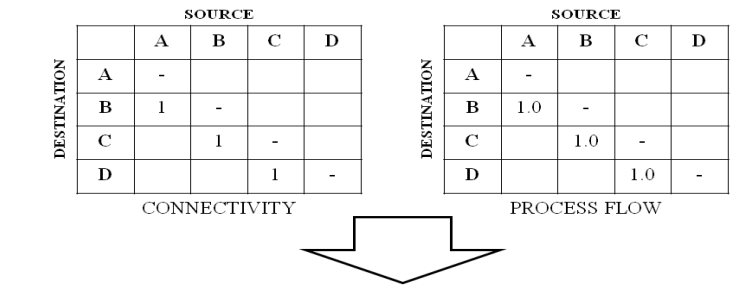

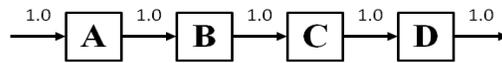

## Mutation, Order 1

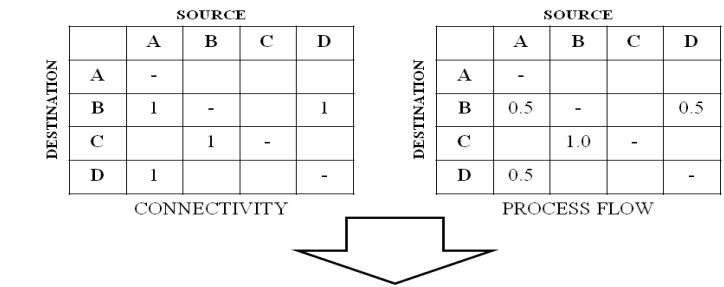

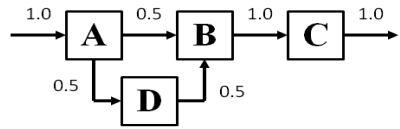

## Mutation, Order 2

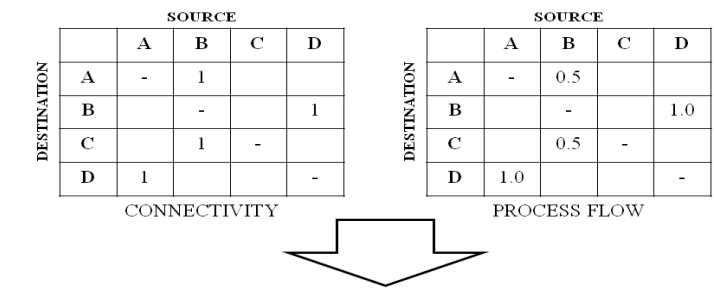

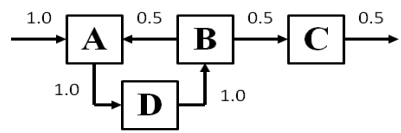

**Figure 7.** The effects of mutations as compared to the initial condition (top) for order 1 (middle and order 2 (bottom) mutational change, respectively.

One way to add this complexity is to hook up a great number of interconnected nodes to the input and output of the biological RGM. Due to its massively interconnected nature, this new construct called a convolution architecture should produce highly nonlinear effects, some of which are expected to carry over into the biological RGM. The connectivity and topology of the I/O complex network determines the level of spaghettification experienced in evolution.

**Network Topologies and the effects of Biological RGMs.** In Figure 8, a toy example of a bipartite network connected to a biological RGM is shown. In the case of a bipartite network, the input and output are distinct and feedback is nonexistent. This gives us a linear input into the biological RGM, which may generally be free of noise but also devoid of transformative or stochastic resonance effects. The bipartite convolution architecture can also be referred to as a grounded network, and results in limited spaghettification. Furthermore, only the input directly affects the evolution of the biological RGM.

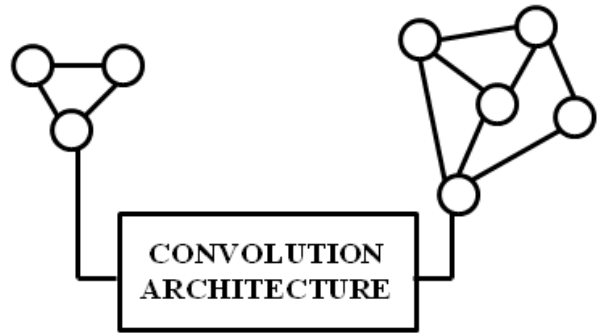

**Figure 8.** A standard bipartite network that serves as the input and output of a biological RGM (convolution architecture). In the case of a bipartite network, there is no feedback and the amount of "spaghettification" is limited.

**Embedded Biological RGMs and Convolution Architectures.** A truly embedded biological RGM is connected to a unipartite complex network. While this network topology might contain modules or clusters independent of the biological RGM, it is connected enough to provide feedback (as a re-entrant input) from the initial biological RGM output. Rather than a simple input, the input in an embedded biological RGM is full of higher-order information and can enable stochastic resonance (e.g. order from noise) effects [26]. In complex networks, stochastic resonance may serve to amplify the long-range transmission of information and enhance information from other genes and circuits in the genetic background [27]. This is the essence of spaghettification: a genetic background that collectively (and indirectly) influences or otherwise supports the evolution and sub-optimal function of a rather simple biological circuit.

## Conclusions

The popular notion of natural selection as "survival of the fittest" may not tell the entire story of evolution. Instead, it might be said that natural selection is also about "survival of the most complex". With that in mind, the biological RGM may serve as an alternative hypothesis to parsimony. In this paper, we have discussed three instances of Rube Goldberg Machine: engineered, mechanical RGMs and two types of biological RGM (model and empirical phenomenon). The mechanical RGM serves as a metaphor for the "maximum number of steps" principle, it might be hard to clearly identify the function of actively evolving RGMs in a biological system. Future experiments might include looking at flows and other mechanical phenomena using microfluidics or other hybrid biological-mechanical models. With the inclusion of biological RGMs into convolution architectures, the effects of massive connectivity can also be used to better understand how so-called spaghettified and highly nonlinear structures can emerge as mechanisms that confer high levels of fitness to the constituent organism. Finally, it is important to keep in mind that convoluted processes can yield relatively straightforward structures just as often as straightforward processes can yield convoluted outputs. In biology, interactions between phenomena such as genes, proteins, and behavior are the norm [28].

## Acknowledgements

I wish to thank participants in the graduate seminar "Evolution of Nervous Systems" and the evolutionary modeling community (e.g. BEACON Center) at Michigan State University for their ideas and discussion. I also wish to thank Rube Goldberg for his hybrid creations.

## Methods

**Biological RGM modeling conventions.** The models presented in Figures 2 and 3 are to be understood using a few conventions which will be presented here. The inputs are introduced at the left hand side of the diagram (arrow leading into element A), while the outputs are expelled on the right (arrow that does not point to any element). Each RGM consists of a set of elements (boxes) and connections (lines) which denote an active process. Lines can have arrows, which specify the direction in which a process is flowing. Mutations can be introduced which modify these connections, either by shortening them (thus rendering their output to a downstream element useless) or changing the direction of their flow (which can render the RGMs final output useless). Finally, it is shown in Figure 3 that elements can be grouped into subsets and connected into banks similar to random connectivity in a neural network.

**Microfludic-inspired instantiations of biological RGMs.** The microfluidic-inspired models of vasculature are shown in Figure 5. Each channel begins with a smooth, non-bifurcating morphology and a Couette flow. At each generation, a mutation or inversion can occur with probability $p_m$ and $p_i$, respectively. Each change to the channel will introduce turbulence to that channel's flow and correspondingly decrease its index of laminarity ($I_{lam}$).

## Parameters

The following parameters describe components of the microfluidic-like model demonstrated in Figure 4.

**Index of Laminarity.** The index of Laminarity ($I_{lam}$) can be defined conceptually as

$$I_{lam} \sim \frac{1}{Ro} \qquad [1]$$

where $R_o$ is the Reynolds number. In this example, higher Reynolds number values translate into less laminarity, and disrupted function.

**Mutation.** Mutations occur as "shifts" in the channel morphology. The location of these shifts and their interval are generated at random, but occurs on the right end, left end, and middle of the channel.

Mutations occur at a rate of µ, and are independent for each channel. The parameter $M_{i,j}$ can be defined as

$$M_{i,j} = \mu(I, P) \qquad [2]$$

where I (1, 2,….n) is the interval over which the mutation occurs, and P (left end, middle, right end) is the position in which the mutation occurs. The mutation is defined as a rate as a function of its physical position on the channel.

**Inversion.** Inversions occur as "flips" in the morphology of a channel, so that the flipped part of the channel is either orthogonal (90 degrees) or diagonal (45 degrees) to the rest of the channel. The location of these flips and their interval are determined at random.

Inversions occur at a rate of λ, and are independent for each channel. The parameter $I_{i,j}$ can be defined as

$$I_{i,j} = \lambda(A, I, P) \qquad [3]$$

where A (45°, 90°) is the angle at which the target position is inverted, I (1, 2,….n) is the interval over which the inversion occurs, and P (left end, middle, right end) is the position in which the inversion occurs. The inversion is defined as a rate as a function of its physical position on the channel.

**Determining convolution in output**
Changes that occur in our biological heuristics can be characterized by comparing the output of single elements versus a serial network of elements. One way this is thought to occur is through convolution of output.

$$RGM_n = O_A * O_B * O_C \qquad [4]$$

where $RGM_n$ is the RGM in question, and O is the output for each element.

RGM output. Output can be defined as a range of behaviors or functions formally defined within a phase space. Mutations and inversions can decrease the heterogeneity of output, particularly desirable and essential functions. When additional features are added, the output space is enlarged and the potential exists for these desirable and/or essential functions to be compensated (for expected relationship, see Figure 6).

Generally, the following relationship is expected to hold (as in Figure 6)

$$(A*B') < (A*B), (A*B*C) \quad\quad [5]$$

In addition, it is generally assumed that the output of multiple elements is synergistic with respect to their output as independent, uncoupled units.